\documentclass[10pt]{article}
\pdfoutput=1
\usepackage{lineno}
\usepackage{scicite}
\usepackage{times}
\usepackage{color}
\usepackage{graphicx,epsfig,epsf}
\usepackage{graphicx}
\usepackage{amssymb}
\usepackage{rotating}
\usepackage[update,prepend]{epstopdf}
\usepackage{ulem}
\usepackage[hang,raggedright]{subfigure}
\usepackage{multicol}
\usepackage{caption}
\usepackage[colorlinks, linkcolor = black, citecolor = black, filecolor = black, urlcolor = blue]{hyperref} 

\newcommand{\msun}{\ifmmode\mbox{M}_{\odot}\else$\mbox{M}_{\odot}$\fi}
\newcommand{\rsun}{\ifmmode\mbox{R}_{\odot}\else$\mbox{R}_{\odot}$\fi}
\newcommand{\psr}{PSR~J1311$-$3430}
\newcommand{\FGLtwo}{2FGL~J1311.7$-$3429}

\newenvironment{sciabstract}{%
\begin{quote} \bf}
{\end{quote}}

\newcounter{lastnote}

\title{Binary Millisecond Pulsar Discovery\\ via Gamma-Ray Pulsations}

\author{
H.~J.~Pletsch$^{1,2\ast}$, 
L.~Guillemot$^{3}$, 
H.~Fehrmann$^{1,2}$, 
B.~Allen$^{1,2,4}$, 
M.~Kramer$^{3,5}$,  
C.~Aulbert$^{1,2}$, \and
M.~Ackermann$^{6}$,
M.~Ajello$^{7}$,
A.~de~Angelis$^{8}$,
W.~B.~Atwood$^{9}$
% ...
}

\date{}

\begin{document} 

\baselineskip24pt  

\maketitle 

\vspace*{-1em}\noindent
\normalsize{$^\ast$To whom correspondence should be addressed. Email: \href{mailto:holger.pletsch@aei.mpg.de}{holger.pletsch@aei.mpg.de}}\\[0.4em]
Affiliations are listed at the end of the paper.\\

\pagebreak

\begin{sciabstract}
Millisecond pulsars, old neutron stars spun-up by accreting matter 
from a companion star, can reach high rotation rates of hundreds of revolutions 
per second. Until now, all such ``recycled'' rotation-powered pulsars have been 
detected by their spin-modulated radio emission.
In a computing-intensive blind search of gamma-ray data from the Fermi Large 
Area Telescope (with partial constraints from optical data), 
we detected a 2.5-millisecond pulsar, \psr. 
This unambiguously explains a formerly unidentified gamma-ray source that 
had been a decade-long enigma, confirming previous conjectures.
The pulsar is in a circular orbit with an orbital 
period of only 93~minutes, the shortest of any spin-powered pulsar binary 
ever found. 
\end{sciabstract}

\begin{multicols}{2}

Almost exactly 30 years ago, radio observations detected the first 
neutron star with a millisecond spin period \cite{Backer+1982}.
Launched in 2008, the Large Area Telescope (LAT) on the 
Fermi Gamma-ray Space Telescope \cite{generalfermilatref} 
confirmed that many radio-detected millisecond pulsars (MSPs)
are also bright gamma-ray emitters~\cite{8MSPs2009}. 
In each case, gamma-ray (0.1 to 100~GeV) pulsations were revealed 
by using rotation parameters obtained from radio telescopes \cite{Smith+2008}
to assign rotational phases to LAT-detected photons.

The Fermi LAT also provides sufficient sensitivity to
detect pulsars via direct searches for periodicity in the sparse
gamma-ray photons.
Such blind searches~\cite{blindsearches} of LAT data for solitary pulsars have so far unveiled 36 younger gamma-ray
pulsars \cite{16gammapuls2009,8gammapuls2010,Pletsch9pulsars,Pletsch+2012-J1838}
with rotation rates between 2 and 20~Hz.
In the radio band, all but four of these objects remain completely undetected 
despite deep follow-up radio searches~\cite{Ray+2012}. 
This is a large fraction of all young gamma-ray emitting neutron stars 
and shows that such blind-search gamma-ray detections are essential 
for understanding the pulsar population \cite{Watters+2011}.
However, no MSP has been detected via gamma-ray pulsations until now, 
and so we have not been able to see whether a similar population of radio-quiet MSPs exists.

The blind-search problem for gamma-ray pulsars is computationally demanding, 
because the relevant pulsar parameters 
are unknown a priori and must be explicitly searched. 
For observation times spanning several years,
this requires a dense grid to cover the multi-dimensional parameter space,
with a tremendous number of points to be individually tested.
Blind searches for MSPs in gamma-ray data are vastly more difficult than for slower 
pulsars largely because the search must extend to much higher spin frequencies
[to and beyond 716~Hz \cite{Hessels+2006}].
Furthermore, most MSPs are in binary systems, where the additionally unknown 
orbital parameters can increase the computational complexity by orders of magnitude.
Thus, blind searches for binary MSPs were hitherto virtually unfeasible.

We have now broken this impasse, detecting a binary MSP, denoted \psr{}, 
in a direct blind search of the formerly unidentified gamma-ray source \FGLtwo, 
one of the brightest listed in the Fermi-LAT Second Source 
Catalog [2FGL \cite{FermiSecondSourceCatalog}]. 
This source also had counterparts in several earlier gamma-ray catalogs and
was first registered in data from the Energetic Gamma Ray Experiment 
Telescope [EGRET \cite{1EGcat1994}] on the Compton Gamma Ray Observatory.

In a search for potential optical counterparts of \FGLtwo{}, 
Romani \cite{Romani2012} identified a quasi-sinusoidally modulated optical flux
with a period of 93~minutes and conjectured this to be a 
``black-widow'' pulsar binary system \cite{Fruchter+1988}.
In this interpretation, an MSP strongly irradiates what is left of the donor 
companion star to eventually evaporate it.
This plausibly explained the observed brightness variation
resulting from strong heating of one side of the companion by the pulsar radiation. 
Associating this optical variation with the orbital period of the putative 
binary system constrained the ranges of orbital search 
parameters and also confined the sky location for the search.
Thus, these constraints made a blind binary-MSP search in LAT data feasible;
however the computational challenge involved remained enormous.
To test the binary-MSP hypothesis 
as the possible nature of \FGLtwo{}, we developed a method to
search the LAT data for pulsations over the entire relevant parameter space.

Under the black-widow interpretation, the search is confined
toward the sky location of the potential optical counterpart 
and the orbit is expected to be circular, leaving a five-dimensional search space.
The individual dimensions are 
spin frequency~$f$, its rate of change~$\dot f$, the orbital period~$P_{\rm orb}$, 
time of ascending node~$T_{\rm asc}$, and $x=a_p\,\sin\iota$,
the projection of the pulsar semi-major axis~$a_p$ onto the line of sight
with orbital inclination angle~$\iota$.
We designed the blind search to maintain sensitivity to very high  
pulsar spin frequencies, $f < 1.4$~kHz, and values of $\dot f$ typical for MSPs, 
$-5\times10^{-13}$\,Hz\,s$^{-1} < \dot f < 0$.
Although the optical data constrain $P_{\rm orb}$ and $T_{\rm asc}$, 
the uncertainties are by far larger than the precision necessary for
a pulsar detection. 
This required us to search ranges of $P_{\rm orb} = 5626.0 \pm 0.1$\,s 
and $T_{\rm asc} = 56009.131\pm 0.012$\,MJD (modified Julian days) 
around the nominal values \cite{Romani2012}, and \mbox{$0 < x < 0.1$~lt-s} (light-seconds).

Searching this five-dimensional parameter space fully coherently
given a multiple-year data time span is computationally impossible.
To solve this problem, we used the hierarchical (three-staged) search strategy 
that previously enabled the detection of 10 solitary, younger (i.e. non-MSP) 
pulsars in blind searches of LAT data \cite{Pletsch9pulsars,Pletsch+2012-J1838},
exploiting methods originally developed to detect gravitational waves
from pulsars \cite{bc2:2000,PletschAllen2009,Pletsch2010,PletschSLCW2011}.
Here we expanded this approach to also search over binary orbital parameters.
The first stage of the hierarchical scheme is the most computing-intensive and
uses an efficient ``semi-coherent'' method \cite{Pletsch9pulsars}, 
extending the method of Atwood et al.\cite{Atwood2006}. This step involves (incoherently) combining 
coherent Fourier power computed using a window of 2$^{20}$\,s ($\sim$12~days) 
by sliding the window over the entire LAT data set (hence the term ``semi-coherent''). 
In a second stage, significant semi-coherent candidates are automatically 
followed up through a fully coherent analysis made possible
because only a small region of parameter space around the candidate is explored.
A third stage further refines coherent 
pulsar candidates by including higher signal harmonics [using 
the $H$-test \cite{deJaeger1989,KerrWeightedH2011}]. 
The computing cost to coherently follow up a single semi-coherent candidate 
is negligible relative to the total cost of the first stage.
Therefore, constructing the search grid of the semi-coherent stage 
as efficiently as possible is of utmost importance.

The key element in constructing an optimally efficient grid for the semi-coherent search
is a distance metric on the search space \cite{bc2:2000,PletschAllen2009,Pletsch2010,Messenger2011}.
The metric provides an analytic geometric tool measuring the expected fractional loss 
in signal-to-noise ratio (S/N) squared for any given pulsar-signal location at a nearby grid point. 
The metric is obtained from a Taylor expansion of the fractional loss to second order 
around the parameter-space location of a given signal.
In contrast to searching for solitary pulsars, a difficulty in the binary case is that the metric 
components explicitly depend on the search parameters \cite{Messenger2011}.                                                                             
Thus, the metric (and so the grid-point density required to not miss a signal) 
changes across orbital parameter space. Constructing a simple lattice with constant 
spacings would be highly inefficient, resulting in either vast over- or under-covering
of large parameter-space regions.
We developed a grid construction algorithm \cite{som} that effectively uses 
the metric formalism. Orbital grid points were first placed at random, 
then those that were either too close together or too far apart according to the metric
were moved (barycentric shifts), minimizing the maximum possible loss 
in S/N for any pulsar signal across the entire search parameter space. 
By design the resulting grid \cite{som} ensured never losing
more than 30\% in S/N for any signal parameters.

The input LAT data we prepared for this search spanned 
almost 4 years (1437~days) and includes gamma-ray photons
with LAT-reconstructed directions within 15$^\circ$ around 
the targeted sky position~\cite{som}.
To improve the S/N of a putative pulsar signal, we assigned 
each photon a weight \cite{KerrWeightedH2011} 
measuring the probability of originating from the conjectured pulsar, 
computed with a spectral likelihood method \cite{som}.
The gamma-ray spectrum of \FGLtwo{} is best modeled by an exponentially cut-off power law (Fig.~S1),
with spectral parameters reminiscent of other gamma-ray pulsars  (Table~1).
The computational work of the search was done on the ATLAS cluster 
in Hannover, Germany. Soon after initiation, the searching procedure convincingly 
detected \psr{}.

Following the blind-search detection, we refined the pulsar 
parameters further in a timing analysis \cite{Ray2011}.
We obtained pulse times of arrival (TOAs) from subdividing the 
LAT data into 40~segments of about equal length. 
We produced a pulse profile for each segment using the initial pulsar parameters, 
and cross-correlated each pulse profile with a multi-Gaussian template derived 
from fitting the entire data set to determine the TOAs.
We used the Tempo2 software~\cite{Tempo2} to fit the TOAs 
to a timing model including sky position, $f$, $\dot f$,
and binary-orbit parameters (Fig.\,1 and Table 1). 
We found no statistically significant evidence for 
orbital eccentricity at the $e < 10^{-3}$ level. We measured a marginal
evidence for a total proper motion of $8 \pm 3$ milliarcseconds/year. 
Generally, the observed value of $\dot f = (-3.198\pm0.002) \times 10^{-15}$\,Hz\,s$^{-1}$ 
is only an upper limit of the intrinsic frequency change~$\dot{f}_{\rm in}$,
because of the Shklovskii effect in which Doppler shifts caused 
by the proper motion can account for part of~$\dot f$. 
Under the assumption that the proper motion of \psr{} is small enough to
approximate $\dot f \cong \dot{f}_{\rm in}$, we 
derived further quantities from the pulsar rotational parameters~(Table~1).

The rotational ephemeris of \psr{} also provides constraints on the companion 
mass~$m_c$ through the binary mass function that combines
$x$, $P_{\rm orb}$, and the gravitational constant~$G$,
\begin{eqnarray}
  f(m_p,m_c) &=& \frac{4\pi^2}{G}\frac{x^3}{P_{orb}^2}
  = \frac{(m_c \, \sin \iota)^3}{(m_c+m_p)^2} \nonumber\\
  &=& (2.995 \pm 0.003) \times 10^{-7} \msun 
\end{eqnarray}
where $m_p$ is the pulsar mass and $\msun$ is the mass of the Sun. 
Typical MSP masses are 1.35 to 2.0$\,\msun$.
Assuming $m_p=1.35\,\msun$ and $\iota=90^\circ$ (orbit is edge-on) yields
the minimum companion mass, $m_c > 8.2\times10^{-3}\,\msun$,
which is only about eight times the mass of Jupiter.
By means of Kepler's third law and typical MSP masses ($m_p\gg m_c$), 
the binary separation, $a=a_p+a_c$
is accurately approximated by $a=0.75\,\rsun (m_p/1.35\,\msun)^{1/3}$,
where $\rsun$ is the radius of the Sun.
Thus, \psr{} is likely the most compact pulsar binary known.

The compact orbit and the optical flaring events \cite{Romani2012}
suggest that the pulsar heating is driving a strong, possibly variable, stellar wind 
of ablated material of the companion. Interactions with the companion wind could 
affect the gamma-ray flux observed. In a dedicated analysis \cite{som},
we found no evidence for a modulation at the orbital period 
of the gamma-ray flux or its spectrum.

We also examined the gamma-ray spectral parameters of  \psr{} 
as a function of rotational phase \cite{som}. 
Dividing the data into 10 segments according to different rotational-phase
intervals, we spectrally analyzed each segment separately.
In line with the background estimation in the pulse profile 
(Fig.~1), we detected significant gamma-ray emission at all phases. 
The gamma-ray spectrum in the off-pulse phase interval (Fig.~S2) 
is better modeled by an exponentially cut-off power law, potentially indicative of magnetospheric origin 
from the pulsar, rather than by a simple power law which would more likely suggest
intrabinary wind shock emission.

Repeated, sensitive radio searches of the previously unidentified gamma-ray source, including
Green Bank Telescope observations at $820$~MHz 
gave no pulsar detection \cite{Ransom2011}.
However, material ablated from the companion by the pulsar irradiation might
obscure radio pulses.  At higher radio frequencies decreased scattering and absorption
resulting in shorter eclipses are observed for other black-widow pulsars \cite{Hessels+2006}.

The optical observations provide evidence for strong heating of the 
pulsar companion that is near filling its Roche lobe \cite{Romani2012}.
With $m_p=1.35\,\msun$ and $\iota=90^\circ$,
the Roche lobe radius of the companion is 
to good approximation \cite{Paczynski1971} $R_L = 0.063\,\rsun$. 
The minimum mean density of the Roche-lobe filling companion directly 
follows from the orbital period \cite{accretionpower}, $\bar{\rho} = 45$\,g\,cm$^{-3}$. 
This is twice the density of the planetary-mass
companion of PSR~J1719$-$1438  \cite{Bailes+2011}. 
One scenario for the formation of that system posits an ultra-compact x-ray binary 
with a He or C degenerate donor transferring mass to the neutron star. 
However, van Haaften et al.~\cite{Haaften+2012} argue that angular momentum losses  through 
gravitational-wave emission are insufficient to reach the low masses and short period 
of the PSR~J1719$-$1438 system within the age of the universe. 
Instead, strong heating to bloat the companion or extra angular momentum loss from 
a companion evaporative wind are required.
An alternative scenario \cite{Benvenuto+2012} proposes that a combination of
angular momentum loss and wind evaporation from an initial companion mass of $2\msun$
in a $0.8$\,day orbit can bring the system to low masses and short orbital periods
in $\sim6$~billion years. Indeed, their scenario produces a good match to the $m_c\sim0.01\msun$,
$P_{\rm orb}\sim 0.065$~days seen for \psr. At this point in the evolution the system
is detached, the companion is He-dominated and irradiation has taken over the
evolution. Presumably continued irradiation can drive the system towards
PSR~J1719$-$1438-type companion masses, or produce an isolated MSP.

The direct detection of an MSP in a blind search of gamma-ray data 
implies that further MSPs, including other extreme binary pulsars,
may exist among the bright, as-yet unidentified 2FGL 
gamma-ray sources [e.g. \cite{Romani+2011,Ackermann+2012}] which are too radio faint 
or obscured by dense companion winds to be found in typical radio searches.

\end{multicols}

\bibliography{ms}

\begin{thebibliography}{10}
\bibitem{blindsearches}
The term ``blind search'' is used to reflect the fact that the pulsar
  parameters are unknown a priori, implying that a wide range of pulsar
  parameters must be explicitly searched.
\end{thebibliography}
\bibliographystyle{Science}
\nocite{FermiVela2}
\nocite{Ackermann+2011}
\textbf{Acknowledgments:}
This work was supported by the Max-Planck-Gesellschaft.
The Fermi LAT Collaboration acknowledges support from several agencies and institutes for both development and the operation of the LAT as well as scientific data analysis. These include NASA and Department of Energy (United States), CEA/Irfu and IN2P3/CNRS (France), ASI and INFN (Italy), MEXT, KEK, and JAXA (Japan), and the K.~A.~Wallenberg Foundation, the Swedish Research Council and the National Space Board (Sweden). Additional support from INAF in Italy and CNES in France for science analysis during the operations phase is also gratefully acknowledged.
Fermi LAT data are available from the Fermi Science Support Center (http://fermi.gsfc.nasa.gov/ssc).

\pagebreak

\begin{table}
\caption*{{\bf Table 1.}
Measured and derived parameters for \psr, with
formal 1$\sigma$ uncertainties in the last digits
(dd, days; hh, hours; mm, minutes; ss, seconds).
Spectral parameters are averages over pulse phase.
}
\begin{center}
\begin{tabular}{ll}
\\
\hline \hline
Parameter & Value \\
\hline \hline
Right ascension (J2000.0) (hh:mm:ss) & $13$:$11$:$45.7242(2)$ \\
Declination (J2000.0) (dd:mm:ss) & $-34$:$30$:$30.350(4)$ \\
Spin frequency, $f$ (Hz) & $390.56839326407(4)$ \\
Frequency derivative, $\dot f$ ($10^{-15}$ Hz s$^{-1}$)  & $-3.198(2)$  \\
Reference time scale                                   & Barycentric Dynamical Time (TDB) \\
Reference time (MJD)                                   & $55266.90789575858$ \\
Orbital period $P_{\rm orb}$ (d)               & $0.0651157335(7)$ \\
Projected pulsar semi-major axis $x$ (lt-s)          & $0.010581(4)$ \\
Time of ascending node $T_{\rm asc}$ (MJD)        & $56009.129454(7)$ \\
Eccentricity $e$                              & $< 0.001$  \\
Data span (MJD)  &  	$54682$ to $56119$ \\
Weighted RMS residual ($\mu$s) & $17$ \\
\hline
{\it Derived Quantities} & \\
\hline
Companion mass $m_{c}$ (\msun)    & $> 0.0082$ \\
Spin-down luminosity $\dot{E}$ (erg s$^{-1}$)          & $4.9 \times 10^{34}$ \\
Characteristic age $\tau_{c}$ (years)         & $1.9 \times 10^9$ \\
Surface magnetic field $B_{\rm S}$ (G)    & $2.3 \times 10^8$ \\
\hline
{\it Gamma-Ray Spectral Parameters} & \\
\hline
Photon index, $\Gamma$ & $1.8(1)$ \\
Cutoff energy, $E_c$ (GeV) & $3.2(4)$ \\
Photon flux above $0.1$ GeV, $F$ ($10^{-8}$ photons cm$^{-2}$ s$^{-1}$) & $9.2(5)$ \\
Energy flux above $0.1$ GeV, $G$ ($10^{-11}$ erg cm$^{-2}$ s$^{-1}$) & $6.2(2)$ \\
\hline \hline
\end{tabular}
\end{center}
\end{table}

\clearpage
\begin{figure*}
\centering
	\hspace{-0.2cm}
		\subfigure
		{\includegraphics[width=7.6cm]{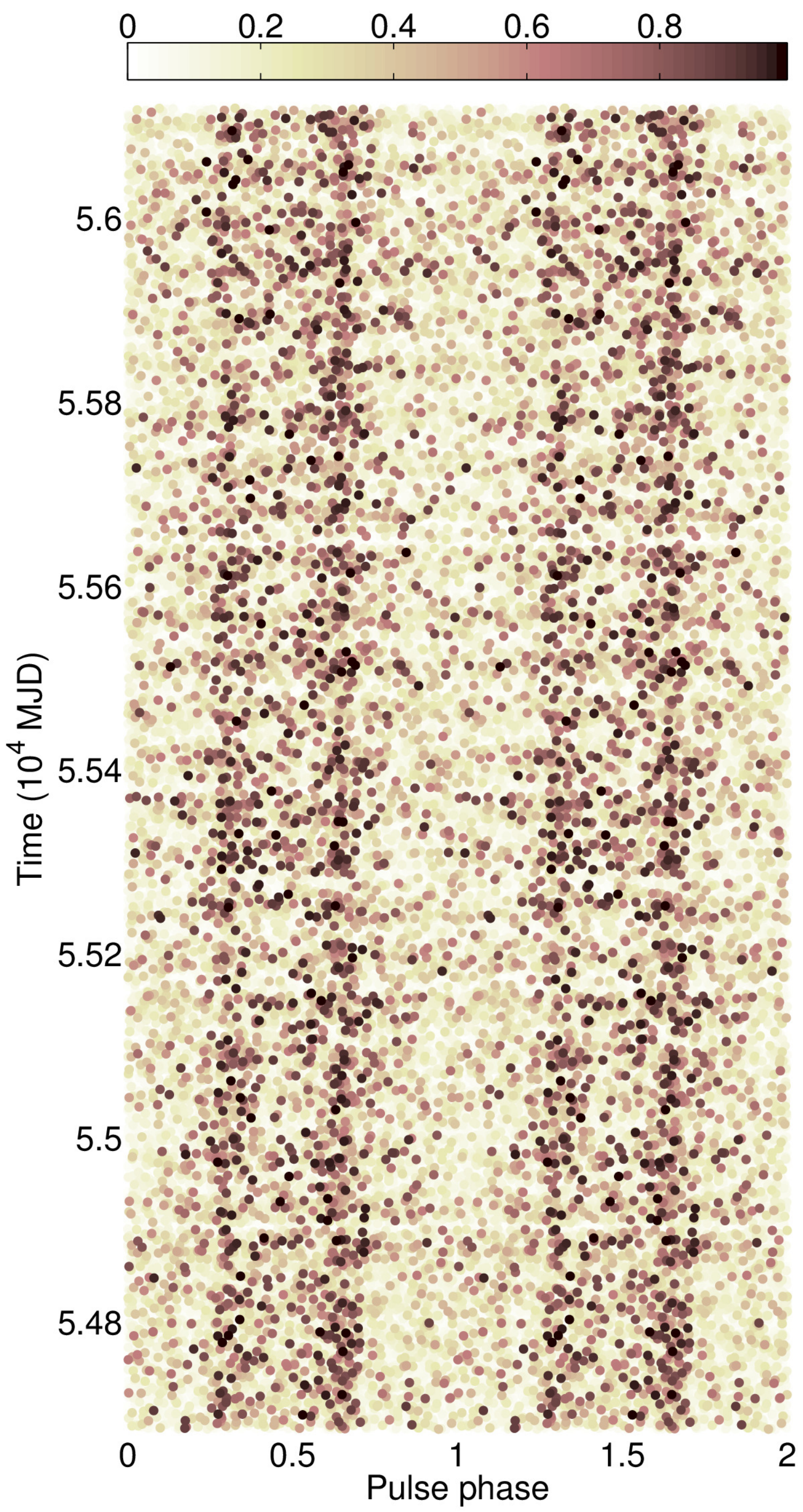}}
		\hspace{0.5cm}
		\subfigure	
		{\includegraphics[width=7.6cm]{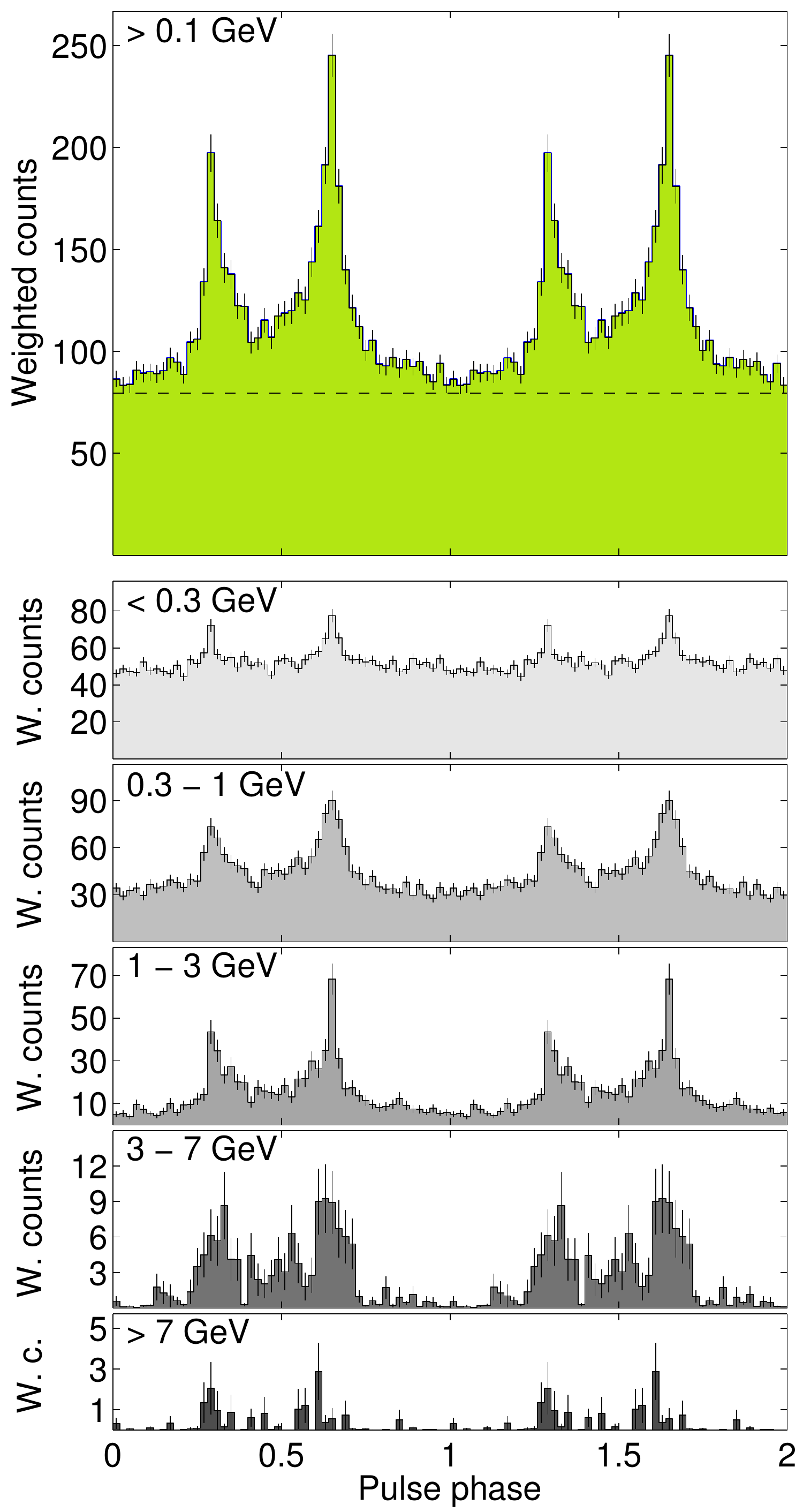}}
	\caption*{{\bf Fig.\,1.}
	Phase-time diagram and gamma-ray pulse profiles for \psr{}.        
	Two pulsar rotations are shown for clarity. 
	({\bf Left}) The pulsar rotational phase for each gamma-ray-photon arrival time;
	probability weights are shown in color code. 
	({\bf Right}) The pulse profiles in different energy bands.
	Each bin is 0.02 in phase and photon weights are used.
         The dashed line indicates the estimated background level from a surrounding annular region.     	          
	\label{f:ph-vs-t-all}}
\end{figure*}

\clearpage

\noindent
Affiliations:\\[1em]
\footnotesize{$^{1}$ Max-Planck-Institut f\"ur Gravitationsphysik (Albert-Einstein-Institut), D-30167 Hannover, Germany}\\
\footnotesize{$^{2}$ Institut f\"ur Gravitationsphysik, Leibniz Universit\"at Hannover, D-30167 Hannover, Germany}  \\
\footnotesize{$^{3}$ Max-Planck-Institut f\"ur Radioastronomie, Auf dem H\"ugel 69, D-53121 Bonn, Germany}  \\
\footnotesize{$^{4}$ Department of Physics, University of Wisconsin-Milwaukee, P.O. Box 413, Milwaukee, WI 53201, USA}  \\
\footnotesize{$^{5}$Jodrell Bank Centre for Astrophysics, School of Physics and Astronomy, The University of Manchester, Manchester M13 9PL, UK}  \\
\footnotesize{$^{6}$  Deutsches Elektronen Synchrotron DESY, D-15738 Zeuthen, Germany }  \\
\footnotesize{$^{7}$  W. W. Hansen Experimental Physics Laboratory, Kavli Institute for Particle Astrophysics and Cosmology, Department of Physics and SLAC National Accelerator Laboratory, Stanford University, Stanford, CA 94305, USA }  \\
\footnotesize{$^{8}$  Dipartimento di Fisica, Universit\`a di Udine and Istituto Nazionale di Fisica Nucleare, Sezione di Trieste, Gruppo Collegato di Udine, I-33100 Udine, Italy }  \\
\footnotesize{$^{9}$  Santa Cruz Institute for Particle Physics, Department of Physics and Department of Astronomy and Astrophysics, University of California at Santa Cruz, Santa Cruz, CA 95064, USA }  \\
\footnotesize{$^{10}$  Universit\`a  di Pisa and Istituto Nazionale di Fisica Nucleare, Sezione di Pisa I-56127 Pisa, Italy }  \\
\footnotesize{$^{11}$  Laboratoire AIM, CEA-IRFU/CNRS/Universit\'e Paris Diderot, Service d'Astrophysique, CEA Saclay, F-91191 Gif sur Yvette, France }  \\
\footnotesize{$^{12}$  Istituto Nazionale di Fisica Nucleare, Sezione di Trieste, I-34127 Trieste, Italy }  \\
\footnotesize{$^{13}$  Dipartimento di Fisica, Universit\`a di Trieste, I-34127 Trieste, Italy }  \\
\footnotesize{$^{14}$  Istituto Nazionale di Fisica Nucleare, Sezione di Padova, I-35131 Padova, Italy }  \\
\footnotesize{$^{15}$  Dipartimento di Fisica e Astronomia "G. Galilei", Universit\`a di Padova, I-35131 Padova, Italy }  \\
\footnotesize{$^{16}$  Istituto Nazionale di Fisica Nucleare, Sezione di Pisa, I-56127 Pisa, Italy }  \\
\footnotesize{$^{17}$  CNRS, IRAP, F-31028 Toulouse cedex 4, France }  \\
\footnotesize{$^{18}$  GAHEC, Universit\'e de Toulouse, UPS-OMP, IRAP, Toulouse, France }  \\
\footnotesize{$^{19}$  Dipartimento di Fisica ``M. Merlin" dell'Universit\`a e del Politecnico di Bari, I-70126 Bari, Italy }  \\
\footnotesize{$^{20}$  Istituto Nazionale di Fisica Nucleare, Sezione di Bari, I-70126 Bari, Italy }  \\
\footnotesize{$^{21}$  Laboratoire Leprince-Ringuet, \'Ecole polytechnique, CNRS/IN2P3, Palaiseau, France }  \\
\footnotesize{$^{22}$  Institut de Ci\`encies de l'Espai (IEEE-CSIC), Campus UAB, 08193 Barcelona, Spain }  \\
\footnotesize{$^{23}$  INAF-Istituto di Astrofisica Spaziale e Fisica Cosmica, I-20133 Milano, Italy }  \\
\footnotesize{$^{24}$  Istituto Nazionale di Fisica Nucleare, Sezione di Perugia, I-06123 Perugia, Italy }  \\
\footnotesize{$^{25}$  Dipartimento di Fisica, Universit\`a degli Studi di Perugia, I-06123 Perugia, Italy }  \\
\footnotesize{$^{26}$  NASA Goddard Space Flight Center, Greenbelt, MD 20771, USA }  \\
\footnotesize{$^{27}$  Center for Research and Exploration in Space Science and Technology (CRESST) and NASA Goddard Space Flight Center, Greenbelt, MD 20771, USA }  \\
\footnotesize{$^{28}$  Department of Physics and Center for Space Sciences and Technology, University of Maryland Baltimore County, Baltimore, MD 21250, USA }  \\
\footnotesize{$^{29}$  National Research Council Research Associate, National Academy of Sciences, Washington, DC 20001, resident at Naval Research Laboratory, Washington, DC 20375, USA }  \\
\footnotesize{$^{30}$  ASI Science Data Center, I-00044 Frascati (Roma), Italy }  \\
\footnotesize{$^{31}$  Laboratoire Univers et Particules de Montpellier, Universit\'e Montpellier 2, CNRS/IN2P3, Montpellier, France }  \\
\footnotesize{$^{32}$  Department of Physics, Stockholm University, AlbaNova, SE-106 91 Stockholm, Sweden }  \\
\footnotesize{$^{33}$  The Oskar Klein Centre for Cosmoparticle Physics, AlbaNova, SE-106 91 Stockholm, Sweden }  \\
\footnotesize{$^{34}$  Agenzia Spaziale Italiana (ASI) Science Data Center, I-00044 Frascati (Roma), Italy }  \\
\footnotesize{$^{35}$  IASF Palermo, I-90146 Palermo, Italy }  \\
\footnotesize{$^{36}$  INAF-Istituto di Astrofisica Spaziale e Fisica Cosmica, I-00133 Roma, Italy }  \\
\footnotesize{$^{37}$  Space Science Division, Naval Research Laboratory, Washington, DC 20375-5352, USA }  \\
\footnotesize{$^{38}$  Universit\'e Bordeaux 1, CNRS/IN2p3, Centre d'\'Etudes Nucl\'eaires de Bordeaux Gradignan, F-33175 Gradignan, France }  \\
\footnotesize{$^{39}$  Department of Physical Sciences, Hiroshima University, Higashi-Hiroshima, Hiroshima 739-8526, Japan }  \\
\footnotesize{$^{40}$  INAF Istituto di Radioastronomia, I-40129 Bologna, Italy }  \\
\footnotesize{$^{41}$  Max-Planck-Institut f\"ur Kernphysik, D-69029 Heidelberg, Germany }  \\
\footnotesize{$^{42}$  Landessternwarte, Universit\"at Heidelberg, K\"onigstuhl, D-69117 Heidelberg, Germany }  \\
\footnotesize{$^{43}$  Department of Astronomy, Graduate School of Science, Kyoto University, Sakyo-ku, Kyoto 606-8502, Japan }  \\
\footnotesize{$^{44}$  School of Physics and Astronomy, University of Southampton, Highfield, Southampton, SO17 1BJ, UK }  \\
\footnotesize{$^{45}$  Centre d'\'Etudes Nucl\'eaires de Bordeaux Gradignan, IN2P3/CNRS, Universit\'e Bordeaux 1, BP120, F-33175 Gradignan Cedex, France }  \\
\footnotesize{$^{46}$  Department of Physics, Center for Cosmology and Astro-Particle Physics, The Ohio State University, Columbus, OH 43210, USA }  \\
\footnotesize{$^{47}$  Science Institute, University of Iceland, IS-107 Reykjavik, Iceland }  \\
\footnotesize{$^{48}$  Department of Physics, Royal Institute of Technology (KTH), AlbaNova, SE-106 91 Stockholm, Sweden }  \\
\footnotesize{$^{49}$  Research Institute for Science and Engineering, Waseda University, 3-4-1, Okubo, Shinjuku, Tokyo 169-8555, Japan }  \\
\footnotesize{$^{50}$  Department of Astronomy, Stockholm University, SE-106 91 Stockholm, Sweden }  \\
\footnotesize{$^{51}$  Istituto Nazionale di Fisica Nucleare, Sezione di Torino, I-10125 Torino, Italy }  \\
\footnotesize{$^{52}$  Department of Physics and Department of Astronomy, University of Maryland, College Park, MD 20742, USA }  \\
\footnotesize{$^{53}$  Hiroshima Astrophysical Science Center, Hiroshima University, Higashi-Hiroshima, Hiroshima 739-8526, Japan }  \\
\footnotesize{$^{54}$  Istituto Nazionale di Fisica Nucleare, Sezione di Roma ``Tor Vergata", I-00133 Roma, Italy }  \\
\footnotesize{$^{55}$  Institute of Space and Astronautical Science, JAXA, 3-1-1 Yoshinodai, Chuo-ku, Sagamihara, Kanagawa 252-5210, Japan }  \\
\footnotesize{$^{56}$  Max-Planck-Institut f\"ur Physik, D-80805 M\"unchen, Germany }  \\
\footnotesize{$^{57}$  Harvard-Smithsonian Center for Astrophysics, Cambridge, MA 02138, USA }  \\
\footnotesize{$^{58}$  Institut f\"ur Astro- und Teilchenphysik and Institut f\"ur Theoretische Physik, Leopold-Franzens-Universit\"at Innsbruck, A-6020 Innsbruck, Austria }  \\
\footnotesize{$^{59}$  NYCB Real-Time Computing Inc., Lattingtown, NY 11560-1025, USA }  \\
\footnotesize{$^{60}$  Department of Chemistry and Physics, Purdue University Calumet, Hammond, IN 46323-2094, USA }  \\
\footnotesize{$^{61}$  Consorzio Interuniversitario per la Fisica Spaziale (CIFS), I-10133 Torino, Italy }  \\
\footnotesize{$^{62}$  Dipartimento di Fisica, Universit\`a di Roma ``Tor Vergata", I-00133 Roma, Italy }  \\
\\

\pagebreak

\normalsize\selectfont

\section*{Supporting Online Material}\vspace*{1em}

\subsection*{Materials and Methods: Fermi-LAT Data Analysis}\vspace*{1em}

\begin{multicols}{2}
\noindent\textbf{LAT data selection}\\
The LAT surveys the entire sky every 3 hours (two orbits). In this work,
we used data taken in this sky-survey mode between 4 August 2008 and 10 July 2012. 
The data were processed using the Fermi 
Science Tools\footnote{http://fermi.gsfc.nasa.gov/ssc/data/analysis/scitools/overview.html}
(v9r28p0). We selected gamma-ray photons belonging to the ``Source'' class under the P7V6  
event selections. Photons with reconstructed zenith angles larger than $100^\circ$ 
were rejected in order to exclude the bright contribution from the Earth's limb. 
We also rejected photons recorded when the instrument was not operating in sky-survey mode, 
or when its rocking angle exceeded $52^\circ$. Finally, we selected only photons
with energies $> 0.1$ GeV and found within $15^\circ$ of the direction of the pulsar.\\[-1em]

\noindent\textbf{LAT data spectral analysis}\\
We determined the gamma-ray spectral properties of \psr{} by performing a binned likelihood 
analysis of this data set, using the \texttt{pyLikelihood} tool. The Galactic diffuse emission 
was modeled using the \textit{gll\_iem\_v02} map cube, 
and the extragalactic emission and the residual instrumental backgrounds were 
modeled jointly using the \textit{iso\_p7v6source} template\footnote{
http://fermi.gsfc.nasa.gov/ssc/data/access/lat/BackgroundModels.html}. The 
spectral model used in this analysis also included the contributions of 2FGL 
sources \cite{FermiSecondSourceCatalog} within $20^\circ$ of the center 
of the field of view, and the contribution from the pulsar was modeled 
using an exponentially cut-off power-law (ECPL) 
of the form $dN / dE \propto E^{-\Gamma} \exp( -E / E_c)$ 
where $E$ represents the photon energy, $\Gamma$ is the photon index and 
$E_c$ is the cutoff energy of the spectrum. The normalizations of the 
diffuse components were left free in the fit, as well as spectral parameters 
of sources within $8^\circ$ of the pulsar sky position. 
The best-fit spectral parameters for \psr{} along with the derived photon 
and energy fluxes are displayed in Table~1 and
characterize the pulse-phase-averaged gamma-ray spectrum of the pulsar (Fig.~S1). 
The measured energy flux is consistent with that published earlier 
for this gamma-ray source \cite{FermiSecondSourceCatalog}. We checked the best-fit
parameters listed in Table~1 with the \texttt{pointlike} 
likelihood analysis tool \cite{KerrWeightedH2011}, and found results that are 
consistent within uncertainties. The tool \texttt{gtsrcprob} and the spectral 
model obtained from the likelihood analysis were finally used to calculate 
the probabilities that the photons originate from \psr{}.

We also performed a spectral analysis resolved in pulse 
phase (rotational phase) for \linebreak\psr{}.
For this, we divided the data set into ten segments according to pulse phase
and  measured the gamma-ray spectrum in each of those segments 
in a binned likelihood analysis assuming an ECPL model for the pulsar (Fig.~S2). 
As observed for other bright gamma-ray pulsars [e.g. \cite{FermiVela2}], 
the spectral properties of \psr{} evolve strongly with rotational phase, suggesting varying 
emission altitudes and curvature radii of the magnetic field lines. In addition, significant 
gamma-ray emission is detected over all rotational phases: selecting phases 
between $0.8$ and $1.2$ we measured a photon flux for \psr{} 
of $(4.6 \pm 0.9) \times 10^{-8}$ photons\,cm$^{-2}$\,s$^{-1}$, and a photon flux 
of $(2.8 \pm 1.6) \times 10^{-8}$ photons\,cm$^{-2}$\,s$^{-1}$ for pulse phases 
between $0.9$ and $1.1$. In the former phase interval, the exponentially cut-off 
power-law model is preferred over a simple power-law spectral shape at the 
$\sim 3.5 \sigma$ significance level, and at $\sim 2.5 \sigma$ for the latter 
phase interval. This potentially indicates a magnetospheric origin for this ``off-pulse'' emission. 
The existence of magnetospheric emission in the off-pulse region of 
the gamma-ray pulse profiles is predicted from theoretical models
under specific geometrical orientations \cite{Ackermann+2011} and can give 
insights into the pulsar emission geometry.

We also conducted an analysis to look for an orbital modulation
of the gamma-ray flux and the spectrum. We subdivided the orbit into 
ten equally spaced bins, and for each subset of data we performed a binned likelihood analysis.
The flux, the spectral index~$\Gamma$, and the energy cutoff~$E_c$ are 
compatible with a constant value all along the orbit. 
Thus, we proceeded to compute the formal 95\% confidence upper limits 
on the amplitude of a sinusoidal modulation and 
obtained $< 2.6 \times 10^{-8}$~photons cm$^{-2}$ s$^{-1}$
for the flux, $< 0.32$ for $\Gamma$, and $< 2.2$~GeV for~$E_c$.
\pagebreak

\noindent\textbf{Pulsar search in LAT data}\\
To correct for the Doppler modulation due to the satellite's motion in the Solar System,
we applied barycenter corrections to the arrival times of the LAT gamma-ray photons
using the JPL DE405 Solar System ephemeris.

Constructing the parameter-space grid for the semi-coherent search 
(the first stage of the hierarchical search scheme)
as efficiently as possible is of utmost importance, because this
stage dominates the overall computing cost.
For this purpose, we developed an algorithm that effectively utilizes 
the metric formalism. The metric provides a geometric tool 
measuring the expected fractional loss in squared signal-to-noise ratio 
for any given pulsar-signal location at a nearby grid point. 
The metric components along the search-space directions of
$f$ and $\dot f$ are constant across the entire space
[see, e.g., \cite{PletschAllen2009,Pletsch2010}].
In contrast, the orbital metric components (in search-space directions 
of $\{P_{\rm orb}, T_{\rm asc}, x\}$) explicitly depend upon the search 
parameters [see, e.g., \cite{Messenger2011}].                                                                          
This implies that the metric (and so the grid-point density required 
to not miss a signal) changes across orbital parameter space.
Therefore, constructing a simple lattice with constant spacings over
these dimensions would be highly inefficient, resulting in either vast 
over- or under-covering of large parameter-space regions.

In contrast, the grid-construction algorithm we developed follows
a more efficient approach. While using constant spacings in 
$f$ and $\dot f$, grid points over $\{P_{\rm orb}, T_{\rm asc}, x\}$
are first placed at random [e.g.,\cite{Harry+2009}].
Then those that are either too close together or too 
far apart according to the metric are moved (barycentric shifts), 
minimizing the maximum possible loss 
in signal-to-noise ratio for any pulsar signal across 
the entire search parameter space. 
To accelerate this otherwise computationally-bound process, we divided the search 
space into sub-volumes to avoid metric distance comparisons between a trial point
and {\it every} other grid point, exploiting an efficient hashing technique. 
For an arbitrary point in parameter space, the index of the enclosing sub-volume 
(and thus the parameters of its neighboring points) is obtained from a hash table
through a fast rounding operation.
By design the resulting grid ensured never losing
more than 30\% of the signal-to-noise ratio for any signal parameters.
Finally, we used simulated pulsar signals to validate this design goal.
This process is also highly accelerated by exploiting the hash table, 
because it provides quick access to the closest grid points around 
any given parameter-space location. This way, the search for each 
simulated pulsar signal is computationally inexpensive, because only the
relevant subset of nearest grid points  around each signal are searched.
 
In the blind search, we obtained
the semi-coherent detection statistic (coherent Fourier power) computed 
using a coherence window of $T = 2^{20}$\,s ($\sim$12~days) being incoherently 
combined by sliding the window over the entire 4~years of LAT data) 
over the entire $f$ grid by exploiting the efficiency of the fast Fourier transform (FFT) algorithm.
For this purpose, we divided total $f$ search range into separate bands 
using a heterodyning bandwidth of $\Delta f_{\rm BW} = 128$\,Hz. Thus, the FFT
contains $T\,\Delta f_{\rm BW} \cong 10^{8}$ frequency bins. 
This choice for $\Delta f_{\rm BW}$ allowed us to fit the computation into memory 
on the ATLAS computing cluster to maximize performance. 
In the $\dot f$ direction, we analyzed about $10^2$ uniformly spaced grid points.
The use of separate frequency bands is also extremely advantageous in view
of the orbital grids, which we adapted to each band.
This further reduced the computational cost, since the total number of required grid 
points to cover the orbital parameter space
$\{P_{\rm orb}, T_{\rm asc}, x\}$ is about
$10^{7} \, \left( f_{\rm max} / 700\,{\rm Hz} \right)^3$ and
increases with $f_{\rm max}$ cubed, 
where $f_{\rm max}$ is the highest spin frequency (most conservative choice) in 
the search band (before heterodyning). Thus, in total the search grid covering the
frequency band near $700$\,Hz comprised about $10^{8}\times10^{2}\times10^{7} = 10^{17}$ points.
We note that the heterodyning step reduces the bandwidth of the data to $128$\,Hz, which is still
much larger than the narrow bandwidth of the signal (that is about $f\times 10^{-4}$,
which is $\sim 0.04$\,Hz for \psr{}) that remains unaffected by this.
The heterodyning does not alter the pulsar signal shape (i.e., the extension 
in parameter space) - it is merely shifted in the frequency domain and so the metric
is unaffected and still depends on $f_{\rm max}$, the original upper frequency.
This $f_{\rm max}^3$-dependency is readily seen from the fact that the amplitude 
of the pulsar-rotational-phase modulation due to the binary motion is proportional
to the spin frequency~$f$. And, the metric components involve products 
of first-order derivatives of the rotational phase with respect to the 
parameters \cite{bc2:2000,PletschAllen2009,Pletsch2010,Messenger2011}.
As the number of grid points is proportional to the square root of the
metric determinant, it therefore increases with $f^3$ (one $f$ contribution 
from each of the three orbital dimensions searched). 

\end{multicols} 

\begin{figure*}
\centering
	\includegraphics[width=0.99\textwidth]{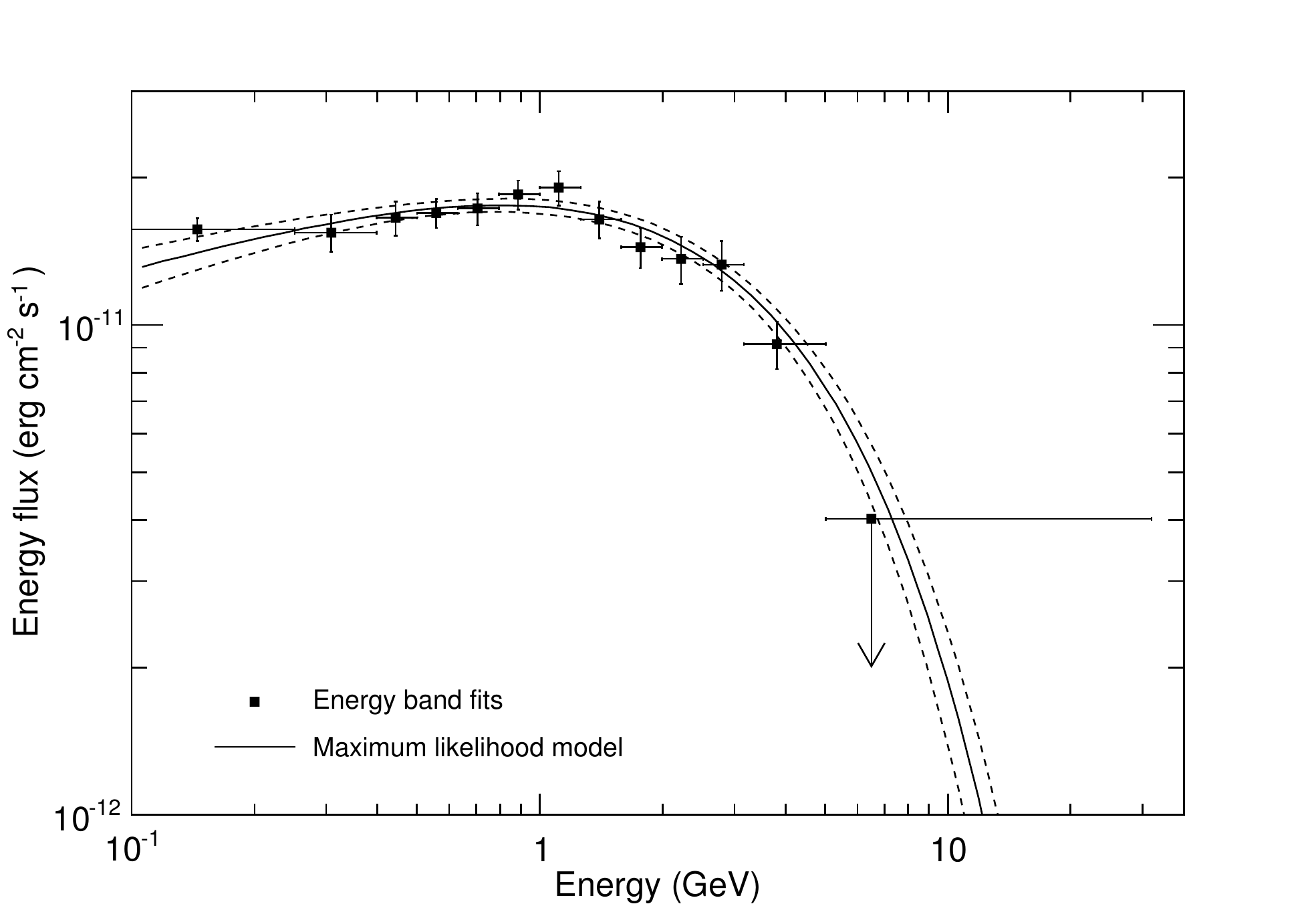}
	\caption*{{\bf Fig.\,S1.}
	 Pulse-phase-averaged gamma-ray spectral energy distribution for \psr{}.
	 Data points (solid squares) are derived from fits of individual energy bands with 
	 variable widths, in which the pulsar is detected with a significance greater than $15\sigma$. 
	 In these individual bands the pulsar emission is modeled with a simple power-law spectrum. 
	 We calculated an upper limit for the highest energy band, in which 
	 the pulsar was not detected with sufficient significance.  
	 The solid curves represents the best-fit model obtained from the likelihood analysis 
	 and dashed curves indicate the $1\sigma$ uncertainties.
	 \label{f:avg_spec} 
	 }
\end{figure*}

\begin{figure*}
\centering
	\includegraphics[width=0.99\textwidth]{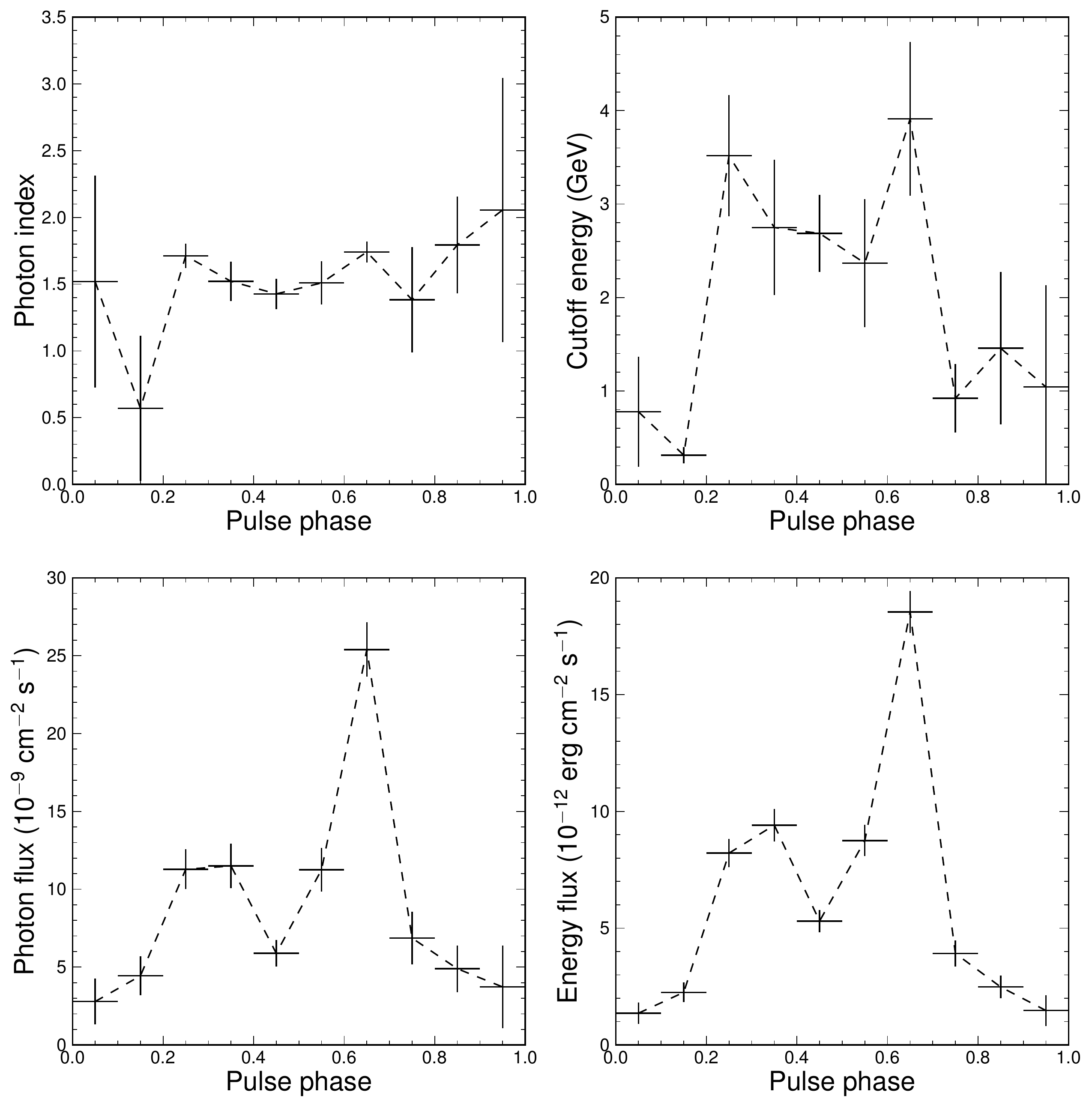}
	\caption*{{\bf Fig.\,S2.}
	 Pulse-phase-resolved gamma-ray  spectral analysis for \psr{}
	 using 10~bins per rotation.
	 The different panels show the
	 photon index ({\bf top left}), 
	 cutoff energy ({\bf top right}), 
	 photon flux above 0.1~GeV ({\bf  bottom left}), and
          energy flux above 0.1~GeV ({\bf  bottom right}).
          Error bars indicate statistical $1\sigma$ uncertainties.
	 \label{f:ph_spec} 
	 }
\end{figure*}

\end{document}